# Unveiling the Dirac feature in the Metallated Carbyne as a Potential Platform for Exploring Widely Separated Majorana Fermions

*C.H Wong*

*Department of Physics, The Hong Kong University of Science and Technology, Hong Kong, China*

**Abstract**

The realization of next-generation quantum computing devices is hindered by the formidable challenge of detecting and manipulating Majorana Fermion in nanomaterials. In this study, we explore a new approach of detecting Majorana Fermion in a metallated carbyne nanowire array. Through comprehensive optimizations, we successfully achieved a local magnetic moment exceeding 3μB, with the average magnetic moment of the entire metallated carbyne surpassing 1μB. Surprisingly, in the absence of spin-orbit coupling, the ferromagnetic Ru metallated carbyne, when coupled with a superconducting Ru substrate, is already able to demonstrate the symmetric opening of a Dirac gap at the gamma point. We discovered that the kink structure of the metallated carbyne plays a crucial role in modulating its topological properties. Moreover, we identified the origin of magnetic hybridization which is intricately linked to the distinctive features found in one-dimensional carbon structures. Our findings not only uncover the unconventional ferromagnetism observed in metallated carbyne but also present an exciting opportunity to realize carbon-based materials capable of hosting Majorana Zero Modes (MZM). This discovery has the potential to further stabilize MZM by decoupling the orbital perturbation from the MZM itself.

**Introduction**

Unlike conventional fermions (e.g. electron), Majorana fermions are zero energy, chargeless, spinless particles under non-Abelian statistics which makes them desire for quantum computing [1-3]. One of the key challenges in developing quantum computers is the preservation of fragile quantum information, known as quantum coherence. Majorana fermions, due to their non-Abelian statistics, are predicted to be more robust against certain types of decoherence caused by interactions with their environment [1-3]. This property makes them potential building blocks for the topological technologies of fault-tolerant quantum computation [4].

Efforts are underway to realize Majorana fermions in condensed matter systems, with three common strategies showing a higher probability of capturing MZMs at the opposite end of the nanowires. These strategies include: (1) reducing the dimensionality of transition-metal dichalcogenides (TMD) with a Dirac point from 2D to 1D through inducing a Dirac gap [5,6,7]; (2) utilizing 1D semiconducting nanowires deposited on an s-wave superconductor under an applied magnetic field [8]; (3) depositing ferromagnetic nanowires on an s-wave superconductor without an applied field [9]. However, it is worth noting that all these methods involve the use of heavy elements due to the requirement of strong spin-orbit coupling for the formation of MZM. For instance, let's consider a monoatomic Fe chain on a Pb(110) substrate [9]. The presence of ferromagnetic exchange splitting leads to the separation of minority and majority bands. If the Fermi level falls within the desired band where only one spin species is occupied, the wavefunction becomes 'spinless' equivalently [8,9]. However, achieving the 'spinless' state always requires the assistance of strong spin-orbit coupling, which modifies the energy levels and opens a Dirac gap in the band diagram [8,9]. This modification enables proximity-induced pairing between the Fe chain and an s-wave Pb superconductor which gives rise to a 1D topological p-wave superconducting state [1-4,8,9]. Within this setup, zero-dimensional MZM emerges at the ends of the Fe nanowire experimentally but it remains a question of whether the separated Majorana particles remain stable if they are widely separated above the micrometer range, as the Fe chain is limited to ~15 atoms only with an irregular atomic alignment in the experiment [9].

The emergence of MZMs in the absence of spin-orbit coupling would be desirable, as it would eliminate the orbital perturbations in the 'spinless state' [8,9]. Carbon-based materials, which exhibit weaker spin-orbit coupling [10] compared to heavy elements, offer a promising avenue in this regard. However, the challenge lies in achieving a strong average magnetic moment and a high Curie temperature in carbon materials [10]. Among carbon materials, nanowires provide a more favorable geometry for the emergence of magnetism. By transforming graphene into zigzag graphene nanoribbons, magnetism can be induced, although the resulting average magnetic moments typically reach only around ~0.2 μB [11]. While falling short of the desired application range of 1μB, the true one-dimensional form of carbon, carbyne, has been investigated for its extraordinary properties. Carbyne, when in isolation, is inherently unstable and typically limited to a length of approximately 50 atoms. However, by utilizing the confinement effect, it becomes possible to extend the length of carbyne to over 1000 atoms [12]. Despite an isolated carbyne not possessing inherent magnetism, the lateral interaction within a closely packed carbyne structure induces an optimal kink configuration, leading to a significant ferromagnetic behavior with a measured Curie point of at least 400K experimentally [13]. But still, the magnetic moments in carbyne remain around 0.2 μB. To overcome this limitation, the introduction of side dopants from Group V elements into the carbyne structure has been proposed. These dopants are expected to maintain the Curie temperature of carbyne close to room temperature, while simultaneously triggering a strong local magnetic moment (~1.5 μB) at the doped sites [14]. Nevertheless, along the carbyne chain, the local magnetic moment of carbon remains relatively low, approximately 0.3μB, and the maximum doping concentration required to generate such magnetism is limited to 15%, which restricts the average magnetic moment of the entire doped carbyne [14].

To overcome these dilemmas, it is imperative to design a long monoatomic nanowire with four desired properties: (1) substantial Curie temperature, (2) strong average magnetic moment, (3) minimal spin-orbit coupling, (4) Dirac feature, in order to study the stability of MZM [1-4,8,9] over extended distances. Metallated carbyne structures, such as -[Cu-C-C]-, have the remarkable capability to facilitate the synthesis of long chains spanning over approximately 1000 regularly aligned atoms [15]. Building upon this characteristic, we propose the engineering of metallated carbyne [15] as a promising avenue for investigating the potential to achieve the four desired properties necessary for capturing MZM. We embark on an investigation into the existence of robust ferromagnetism and the Dirac gap, free from the effects of spin-orbit coupling, within the long metallated carbyne structure when placed on a suitable s-wave superconductor.

**Computational details**

We use CASTEP to search for robust magnetism in metallated carbyne. Geometric optimization is calculated at the spin-unrestricted LDA level [16-18]. Spin-unrestricted GGA-PBE functional is used to simulate the electronic and magnetic properties (unless otherwise specified). The SCF tolerance is $1 \times 10^{-5}$ eV, and the interval of the k-space is 0.0025(1/Å). The maximum SCF cycle is 3000 [16-18]. Norm-conserving pseudopotential is assigned regardless of whether spin-orbital coupling is activated [16-18]. To commence our analysis, we initiate a comparison of the ground-state energies between other metallated carbyne structures and the experimentally attainable Cu metallated carbyne [15]. By evaluating the ground-state energy relative to the Cu metallated carbyne [15], we can gain insights into the relative stability and potential for achieving comparable or even longer chain lengths in the proposed metallated carbyne configurations. After identifying the optimal ferromagnetic metallated carbyne configuration, our next step involves depositing the ferromagnetic chain onto a superconductor substrate and, most importantly, verifying the existence of a Dirac gap [8] without spin-orbital coupling.

**Results**

We present a comparative analysis of Pt-metallated carbyne chains and Cu-metallated carbyne chains [15]. The repeated unit of the infinitely-long Pt-metallated carbyne chain is represented as -[Pt-C-C]-, with a chain-to-chain distance of ~1 nm as shown in the inset of Figure 1. Our investigation reveals that Pt-metallated carbyne chains exhibit a ground-state energy approximately three times more negative than Cu-metallated carbyne chains. We anticipate that the successful fabrication of Pt-metalated carbyne chains will result in longer chains compared to Cu-metallated carbyne chains [15]. In addition, we observe the differential spin density states at the Fermi level with a clear sign of s-p-d hybridization in Figure 1, where p-orbital ferromagnetism dominating. However, the magnetic moments of Pt and C atoms are $0.45\mu_B$ and $0.2\mu_B$, respectively, with the Curie transition temperature $T_{Curie}$ of 7K. By replacing the Pt atom with Rh, Ru, Mo, and Tc, respectively, we observe the presence of finite differential spin density states at the Fermi level in the corresponding metallated carbyne chains: -[Rh-C-C]-, -[Ru-C-C]-, -[Mo-C-C]-, and -[Tc-C-C]-. The differential spin density state at the Fermi level of the Rh metallated chain is the lowest among the investigated chains as illustrated in Figure 2. The magnetic moments of both Rh and C atoms are significantly below $1\mu_B$.

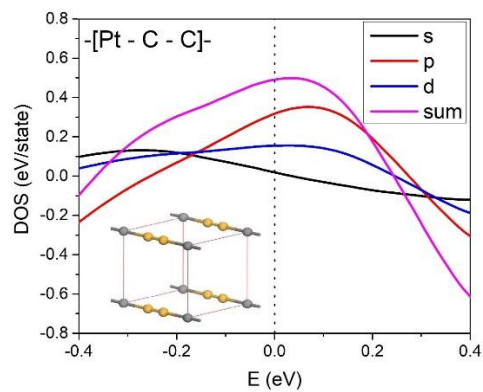

Figure 1: Differential spin density of states (per atom) of Pt-metalated carbyne chain is plotted. The repeated unit of an infinitely-long Pt-metallated carbyne chain is written as -[Pt-C-C]-. The chain-to-chain distance is ~1nm. The inset shows the lattice structure after geometric relaxation. The -[Pt-C-C]- chains are linear. The yellow balls are carbon atoms and the grey balls are Pt atoms. The C-C bond length is 1.27Å. The C-Pt bond length is 2.02Å. The dash line refers to the shifted Fermi level.

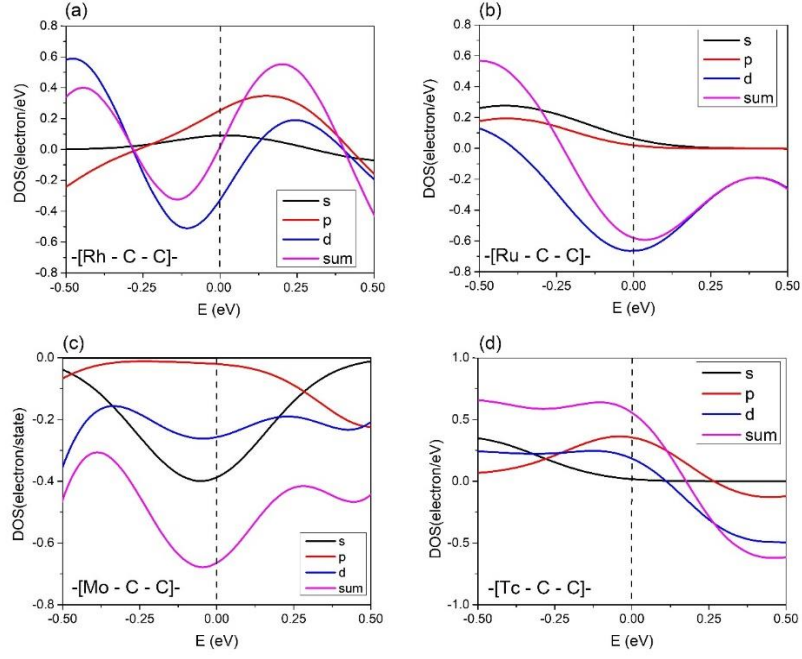

Figure 2: The differential spin density of states per atom of –[X-C-C]- chain (X: Rh, Ru, Mo, Tc) are shown, where the chain-to-chain distance is ~1nm. **a** The C-C bond length is 1.281 Å. The C-Rh bond length is 1.952 Å; **b**, The C-C bond length is 1.286 Å. The C-Ru bond length is 1.999Å; **c** The C-C bond length is 1.271Å. The C-Mo bond length is 2.048Å; and **d** The C-C bond length is 1.273 Å. The C-Tc bond length is 2.046Å, respectively. All –[X-C-C]- chains (X: Rh, Ru, Mo, Tc) are linear.

However, the -[Ru-C-C]-, -[Mo-C-C]-, and -[Tc-C-C]- chains exhibit remarkably large local and average magnetic moments, as indicated in Table 1. In fact, the local magnetic moments of all these metallic atoms surpass 3μB, while the average magnetization along the metallated carbyne exceeds 1 μB. The magnetic moments, Curie temperatures, dominated hybridization and electronic configurations of different metallated carbyne are listed in Table 1. The dominated hybridization of -[Rh-C-C]-, -[Ru-C-C]-, -[Mo-C-C]- and -[Tc-C-C]- are d-orbital, d-orbital, s-orbital, and p-orbital, respectively. On the other hand, we have found that the -[Cu-C-C]-, -[Ag-C-C]-, -[Au-C-C]-, -[Nb-C-C]-, -[Zr-C-C]-, and -[Pd-C-C]- chains can be considered as non-magnetic. These observations indicate that the introduction of Ru, Mo, and Tc atoms into the carbyne chain leads to significant magnetic behavior, characterized by large local and average magnetic moments. As the lateral spacing decreases, the magnetic moments of these chains exhibit a tiny increase in ~1%. Table 2 provides a comprehensive analysis of the spatial dependence of magnetic moments, with a specific focus on the Ru case for clarity reason. Another observation is that for those the metallated carbyne showing magnetism, they exhibit ferromagnetic nature where the Curie temperature is primarily proportional to the magnetic moment of the substitutional dopant X.

Table 1: The magnetism and electronic configurations in -[X-C-C]- chain, where M is the local magnetic moments of the substitutional dopant (X) and carbon (C), <M> is the average magnetic moment per unit cell.

| X | Electron Shells | M: X ($\mu_B$) | M: C ($\mu_B$) | <M> ($\mu_B$) | $T_{Curie}$ (K) | Dominated hybridization |
|---|---|---|---|---|---|---|
| Cu | [Ar] $3d^{10} 4s^1$ | 0.06 ~ 0 | 0.03 ~ 0 | 0 | - | - |
| Ag | [Kr] $4d^{10} 5s^1$ | 0 | 0 | 0 | - | - |
| Au | [Xe] $4f^{14} 5d^{10} 6s^1$ | 0 | 0 | 0 | - | - |
| Pt | [Xe] $4f^{14} 5d^9 6s^1$ | 0.45 | 0.21 | 0.29 | 7 | p-orbital |
| Pd | [Kr] $4d^{10}$ | 0.09 ~ 0 | 0.07 ~ 0 | 0 | - | - |
| Rh | [Kr] $4d^8 5s^1$ | 0.48 | 0.19 | 0.28 | 2 | d-orbital |
| Ru | [Kr] $4d^7 5s^1$ | 3.28 | 0.39 | 1.35 | 140 | d-orbital |
| Tc | [Kr] $4d^5 5s^2$ | 3.5 | 0.25 | 1.33 | 92 | p-orbital |
| Mo | [Kr] $4d^5 5s^1$ | 4.1 | 0.21 | 1.51 | 181 | s-orbital |
| Nb | [Kr] $4d^4 5s^1$ | 0 | 0 | 0 | - | - |
| Zr | [Kr] $4d^2 5s^2$ | 0 | 0 | 0 | - | - |

Table 2: The magnetic moments of -[Ru-$C_2$]- chains in various lateral separations

| Lateral spacing of -[Ru-C-C]- chains (Å) | Magnetic moment of Ru ($\mu_B$) | Magnetic moment of C ($\mu_B$) |
|---|---|---|
| 10 | 3.28 | 0.35 |
| 5 | 3.31 | 0.38 |
| 3 | 3.34 | 0.39 |

In experimental studies of the Cu-metalated carbyne -[Cu-C-C]- chain deposited on the Cu(110) substrate, successful fabrication typically necessitates the use of a substrate mainly composed of the same element [15]. Taking inspiration from this strategy, we adopt a similar approach to relax the Ru-metalated carbyne chain on the Ru substrate, specifically along the [1100] direction. According to our ab-initio simulation, we predict that the presence of a Ru substrate induces the formation of kink structures in the -[Ru-C-C]- chain. The relaxed composite structure, illustrating these kink structures, is depicted schematically in the inset of Figure 3a. Furthermore, we provide a closer view of the kink structure, as shown in the inset of Figure 3b. The average magnetic moments of the -[Ru-C-C]- / Ru[1100] substrate is 1.02$\mu_B$ in which the local magnetic moments of Ru and two C atoms are 2.61$\mu_B$ and 0.22$\mu_B$ and 0.25$\mu_B$, respectively. Upon obtaining the relaxed structure of the -[Ru-C-C]- / Ru[1100] substrate, we observed that the Curie temperature remains approximately 100K. To further investigate the electronic properties, we removed the Ru substrate and computed the band structure of the kink-structured Ru-metallated carbyne alone. Interestingly, in the band structure (Figure 3a), we clearly observe the presence of a Dirac feature [1-3,8,9] at the Gamma point (G) without spin-orbital coupling. Upon closer examination of the band structure, we observe a distinct Dirac gap opening at approximately 4meV in Figure 3b. The opening of the Dirac gap in the band structure is consistently observed across various functional, including PBE, PW91, and WC, etc [16-18], where the magnitude of the Dirac gap ranges from approximately 4-9meV. These consistent observations across different functional in Table 3 suggest that the emergence of the Dirac point is not strongly dependent on the choice of functional.

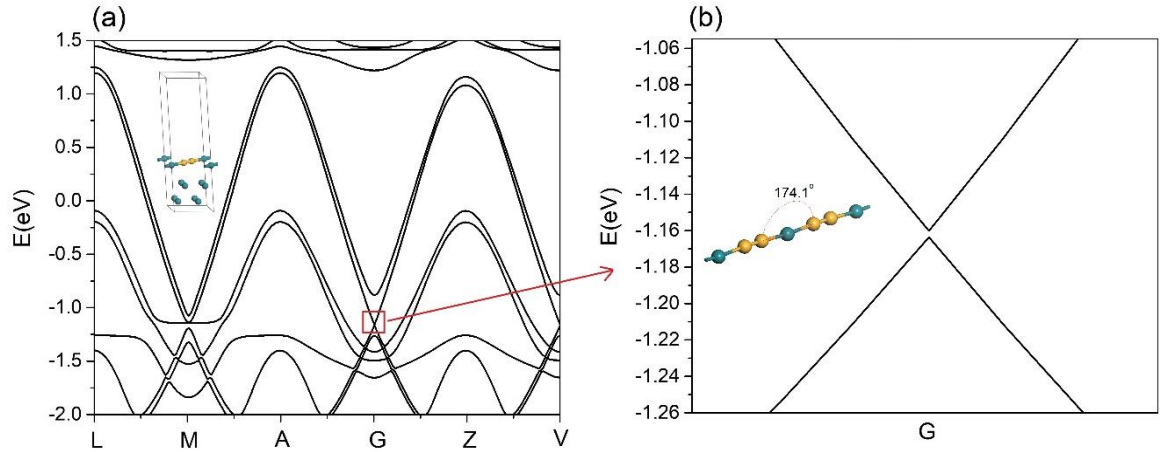

Figure 3: The parallel Ru metallated carbyne chains are deposited on Ru substrate along [1100] direction. After geometric optimization, kink structure appears in the Ru metallated carbyne chain. The -[Ru-C-C]- chains are laterally separated by ~4.6Å. **a** The band diagram of kink structuring Ru metallated carbyne is computed without the Ru substrate. The repeating unit is drawn. The yellow and green balls refer to C and Ru atoms, respectively. **b** The opening of the Dirac point at the gamma point G is observed in -[Ru-C-C]- under a Ru[1100] substrate. The bond distance between the carbon atoms is 1.319Å. The bond distance between the carbon atom and ruthenium is 1.994 Å.

Table 3: Dirac gap and relative position in kink-structured Ru-metallated carbyne

| DFT Functional | Hybridization Gap at the Dirac cone (meV) | Location relative to $E_F$(eV) |
|---|---|---|
| GGA-PBE | 4.3meV | -1.120 |
| GGA-PW91 | 9.8meV | -1.134 |
| GGA-WC | 6.8meV | -1.176 |

## Discussions

The electron configurations of Cu, Ag, Au atoms are [Ar] $3d^{10} 4s^1$, [Kr] $4d^{10} 5s^1$, and [Xe] $4f^{14} 5d^{10} 6s^1$, respectively. However, it is noteworthy that none of them exhibit any noticeable magnetism. Our initial observation suggests that the metalated carbyne fails to exhibit magnetism when the surrounding electrons follow the pattern $nd^i(n+1)s^1$, where i = 10 (or fully filled) and n is principal quantum number. This trend is evident in the metallic cases such as Cu, Ag, Au, where no noticeable magnetic interaction with carbon atoms is observed. For the purpose of conducting a fair comparison in inducing magnetism through carbon atoms, we have selected the elements X: Pd, Rh, Ru, Tc, Mo, Nb, and Zr for building metallated carbyne (Table 1). These elements are chosen because their 4d and 5s electron shells consistently extend beyond the size of a Kr atom. Again, the absence of magnetism in the -[Pd-C-C]- system provides additional evidence supporting the notion that configurations with i = 10 in the $nd^i(n+1)s^1$ arrangement gives a rare magnetism.

Moreover, we observe that the inclusion of substituents Zr (i=2) and Nb (i=4) in the metallated carbyne structure also fails to induce magnetism. This observation, along with the previous findings of non-magnetism in configurations with i=2, 4, and 10 in the surrounding-electron arrangement, i.e, $nd^i(n+1)s^1$, raises the question of whether even values of i are incapable of generating magnetism in X atom via interacting with p-orbital wavefunction in carbon atoms. To address this question, we have conducted a study on the magnetism of the -[Rh-C-C]- system, where i=8. Interestingly, we have observed magnetic moment in this case despite bulk Rh is non-magnetic. This finding suggests that the presence or absence of magnetism in metallated carbyne systems is unlikely attributed to the parity (odd or even) of the value of I. We further investigated the magnetism of -[Mo-C-C]- (i=5) and -[Ru-C-C]- (i=7) systems. In the case of -[Mo-C-C]-, even though the magnetic moment of bulk Mo is very rare, we observed that -[Mo-C-C]- exhibits a strong magnetic moment on the metallic atom, which can be attributed to the fact that a half-filled d shell possesses the highest number of unpaired electrons, thereby triggering to a significant magnetic interaction in the presence of carbon atoms. On the other hand, -[Ru-C-C]- with i=7 displayed an intermediate magnetic moment in Ru atom, falling between that of -[Mo-C-C]- and -[Rh-C-C]-. This is due to the fact that the number of unpaired electrons in the d shell for i=7 is lower than that for i=5, resulting in a reduced magnetic response.

Does the electron configurations of the 5s shell impact the magnetic moment in metallated carbyne? To investigate this, we compare -[Mo-C-C]- and -[Tc-C-C]-, both of which possess a $4d^5$ electron configuration but interact with $5s^1$ and $5s^2$ shells, respectively. Our observations reveal that -[Mo-C-C]- displays a stronger magnetic moment on the metallic atom. This suggests that the presence of unpaired electrons in the 5s shell is more effective in triggering magnetism in these systems. Our study provides confirmation that the presence of unpaired electrons in the d shell alone is not enough to generate magnetism in metallated carbyne systems. This observation is supported by the cases of -[Nb-C-C]- and -[Zr-C-C]-, where unpaired electrons do exist in the d shell, yet magnetism is not observed. Hence, our findings indicate that in order to induce magnetism in -[X-C-C]-, it is necessary for X to achieve at least a half-filled state in the outermost d shell and at the same time it is not in a fully filled state. Otherwise, triggering magnetic interaction between the X atom and C atoms are not effective at all.

Once again, the same reasoning applies to explain the weak magnetic response observed in -[Pt-C-C]. Despite having an electron configuration of $[Xe]4f^{14}5d^96s^1$, where the 5d shell of platinum is not fully filled but exceeds the half-filled state, -[Pt-C-C] still exhibits magnetism. However, due to the presence of only one unpaired electron within the $5d^9$ configuration, the trigger of strong magnetic moment in Pt metallated carbyne is ineffective. Apart from these, we have observed that the dominant type of magnetism in -[X-C-C]- can vary depending on the size of the X atom. In the case of -[Rh-C-C], -[Ru-C-C], -[Tc-C-C], and -[Mo-C-C], with atomic numbers of 45, 44, 43, and 42 respectively, we have noticed a trend: As the size of the X atom increases, there is a transition in the dominant type of magnetism from s-orbital to p-orbital, and finally to d-orbital magnetism.

The choice of utilizing Ru metallated carbyne on a Ru substrate as the experimental system for investigating Majorana zero modes (MZMs) in a monoatomic ferromagnetic nanowire on top of an s-wave superconductor is supported by several scientific reasons [8,9]. The magnetic properties of Ru metallated carbyne are significant in creating the necessary conditions for observing MZMs. While both the monoatomic Fe nanowire and the Ru metallated carbyne exhibit prominent magnetism originating from their d-orbitals, the Ru metallated carbyne exhibits an average magnetic moment of ~1μB, comparable to most transition elements, and a maximum magnetic moment of ~2.6μB, even stronger than bulk Fe, Furthermore, ruthenium, being an s-wave superconductor with a Tc of approximately 0.8K, plays a crucial role in the potential formation of MZM because its Tc is much lower than the $T_{curie}$ of the -[Ru-C-C]- chain (~100K), which is suitable for experimental investigations using a He$^3$ cryostat. Apart from these, the fabrication process of Ru metallated carbyne involves a bottom-up synthesis approach [15], allowing for the positioning of abundant Ru atoms on the top surface of the Ru substrate. By thermal excitation, the Ru atoms escape from the top surface and bond with carbon atoms from carbon-based precursors inside the chamber [15]. Compared to the Cu metallated carbyne [15], the ground state energy of Ru metallated carbyne is approximately 30% more negative, enabling the fabrication of longer monoatomic structures suitable for studying widely separated MZM. In addition, the magnetism in Ru metallated carbyne is not sensitive to lateral spacing, providing flexibility in the fabrication process because any unexpected chain-to-chain distance does not compromise magnetism. But we highly recommend a closely packed Ru-metallated carbyne structure because thermal and quantum phase fluctuations exist in 1D superconducting nanowire which hinder the achievement of zero electric resistance [19-22]. The good regularity of Ru metallated chains is also expected to be superior to that of Fe monoatomic chains, addressing concerns regarding the effect of irregular exchange coupling along the ferromagnetic nanowire on the stability of MZM. Most importantly, the formation of a tiny Dirac gap at the gamma point, approximately a few meV, is crucial for hosting p-wave pairing in the ferromagnetic nanowire and the emergence of MZMs. The Ru metallated carbyne on a Ru substrate is a potential system for achieving this, as the formation of the Dirac gap (~5meV) is sensitive to the kink structure even without spin-orbital coupling. In contrast, the linear Ru metallated carbyne without Ru substrate does not exhibit a Dirac feature in its band structure.

**Conclusion**

This research significantly contributes to the exploration of novel scientific phenomena and holds tremendous potential for advancing the field of quantum computing without relying on spin-orbital coupling. Through a comprehensive analysis of diverse transitional metallic dopants, we have successfully elucidated the fundamental factors that underlie the extraordinary magnetism observed in metallated carbyne. Specifically, the utilization of Ru metallated carbyne on a Ru substrate presents an advanced experimental platform for studying widely separated Majorana zero modes (MZMs). This can be attributed to several advantageous characteristics, including its favorable ferromagnetic properties, the s-wave superconducting nature of ruthenium, the synthesis of monoatomic long chains, improved uniformity in exchange coupling, magnetic insensitivity to lateral spacing, and the presence of a Dirac gap without spin-orbital coupling. This combination of features makes Ru metallated carbyne a desire system for observing and studying MZM.

**Acknowledgment:**

The author appreciates the Department of Industrial and System Engineering at The Hong Kong Polytechnic University to provide simulation support.